\documentclass[10pt,aps,prd,twocolumn,superscriptaddress,nofootinbib]{revtex4-2}

\usepackage{color}
\usepackage{amsmath}
\usepackage{graphicx}
\usepackage{slashed}
\defcitealias{2025PhRvL.134o1002C}{C25}

\usepackage{hyperref}
\hypersetup{
    colorlinks=true,
    linkcolor=blue,
    filecolor=magenta,      
    urlcolor=cyan,
    citecolor=blue
}

\begin{document}

\title{Verifying the failing supernova constraint on dark photons with two-dimensional hydrodynamic simulations}

\author{Kanji Mori}
\email[]{kanji.mori@rk.phys.keio.ac.jp}
\affiliation{Department of Physics, Faculty of Science and Technology, Keio University, 3-14-1 Hiyoshi, Kohoku-ku, Yokohama, Kanagawa 223-8522 Japan}
\author{Tomoya Takiwaki}
\affiliation{National Astronomical Observatory of Japan, 2-21-1 Osawa, Mitaka, Tokyo 181-8588, Japan}
\author{Kazunori Kohri}
\affiliation{National Astronomical Observatory of Japan, 2-21-1 Osawa, Mitaka, Tokyo 181-8588, Japan}
\affiliation{Theory Center, Institute of Particle and Nuclear Studies, KEK, 1-1 Oho, Tsukuba, Ibaraki 305-0801, Japan}
\affiliation{Kavli Institute for the Physics and Mathematics of the Universe (WPI), The University of Tokyo, 5-1-5 Kashiwanoha, Kashiwa, Chiba 277-8583, Japan}

\date{\today}

\begin{abstract}
Recent studies on the dark photon (DP) production in collapsing stars argue that the cooling effect induced by DPs can hinder supernova explosions and lead to a ``failing supernova" constraint on the photon-DP mixing parameter $\epsilon$. In order to verify the idea, we perform two-dimensional neutrino-radiation hydrodynamic simulations coupled with the DP production with the masses of 0.3 and 0.45\,MeV. We find that the shock revival does not happen until the end of the simulations when $\epsilon\gtrsim3\times10^{-9}$. The photon-DP mixing parameter above this value can be excluded by the failing supernova argument. Interestingly, our constraint roughly coincides with the one reported by the previous studies which adopted the post-processing framework. This result motivates one to investigate a wider parameter range of DPs with self-consistent simulations and evaluate uncertainties in the constraint.
\end{abstract}

\maketitle

\section{Introduction}
Dark photons (DPs) are a hypothetical vector boson of an extra $U(1)$ symmetry \cite{Fayet:1980rr,1982JETP...56..502O,1986PhLB..166..196H}. Although DPs are not charged under the Standard Model gauge group, they can interact with photons through the kinetic mixing which leads to the photon-DP oscillation. 

A striking feature of DPs is that they are a candidate of dark matter \cite{2011PhRvD..84j3501N,2012JCAP...06..013A,2016PhRvD..93j3520G,2020JHEP...12..170N,2025arXiv250908932A}. This has motivated physicists to search for the particles with experiments and astrophysical arguments \cite[e.g.][]{2020arXiv200501515F,2021PhRvD.104i5029C,2025arXiv251115785C}. For example, dark matter direct detection experiments such as XENONnT \cite{2022PhRvL.129p1805A}, XMASS \cite{2018PhLB..787..153A}, GERDA \cite{2020PhRvL.125a1801A}, and Majorana Demonstrator \cite{2017PhRvL.118p1801A} provided stringent constraints on the mixing parameter for DPs with the mass of $m_{\gamma'}\sim100$\,keV. Also, DPs can work as an additional cooling source in stars. In fact, comparison between stellar models and globular clusters led to constraints for DPs with $m_{\gamma'}\sim10$--100\,keV \cite{2013PhLB..725..190A,2024JCAP...05..099D}. It is notable that MeV-mass DPs can affect massive stellar evolution as well \cite{2021ApJ...912...13S}.

Core-collapse supernovae are  another useful laboratory for the DP search. For example, Refs.~\cite{2009PhRvD..80g5018B,2016PhRvC..94d5805R,2017JHEP...02..033H,2017JHEP...01..107C,2025JCAP...01..061F} compared the DP and neutrino luminosities from a supernova event. This comparison leads to a constraint because the former should be smaller than the latter  to explain the SN 1987A neutrino burst. When DPs are heavier than $\sim1$\,MeV, they can decay into an electron-positron pair. The decay channel offers constraints based on the Galactic 511\,keV $\gamma$-ray  and the prompt $\gamma$-ray from supernova events \cite{2015NuPhB.890...17K,2019JHEP...02..171D,2022PhRvD.105f3026C,2023PhRvD.108e5014S,2023PhRvD.108e5014S,2025PhRvD.111h3053B}, if DPs can escape from the star. If the decay length of DPs is shorter than the stellar radius, they can play a role in the energy transport in the star and enhance the explosion energy. This argument leads to a calorimetric constraint \cite{2019PhRvD..99l1305S,2022PhRvL.128v1103C}. In addition, when DPs are heavier than $\sim200$\,MeV, their decay may result in high-energy neutrino signals from a nearby supernova event \cite{2024PhRvD.110k5043S}.

Recently, \citet[][hereafter C25]{2025PhRvL.134o1002C} proposed an idea to constrain the DP mass and the mixing parameter on the basis of supernova explodability. When $m_{\gamma'}\sim0.1$\,MeV, DPs can be actively produced in the gain region through a resonance. In this case, the additional cooling counteracts the neutrino heating and hinders the supernova explosions. Since at least some of collapsing stars clearly explode in reality, a ``failing supernova" constraint can be obtained, which supersedes the traditional supernova cooling limit in the mass range $m_{\gamma'}\sim0.1$--0.4\,MeV. A similar argument has been applied to light sterile neutrinos \cite{2014PhRvD..89f1303W,2025PhRvD.111h3046M}, which can be produced in the gain region through the matter-induced neutrino oscillation. 

\citetalias{2025PhRvL.134o1002C} performed post-processing calculations to estimate the DP production and assumed that the supernova becomes failed if the DP cooling rate integrated over the gain region is higher than 20\% of the neutrino cooling rate.  However, recent studies on the supernova mechanism indicate that explodability depends on the progenitor in a complicated way. Although some criteria that determine explodability from the progenitor structure  have been proposed \cite{2011ApJ...730...70O,2012ApJ...746..106P,2016ApJ...818..124E,2016ApJ...821...38S,2022MNRAS.515.1610G,2023ApJ...949...17B,2025A&A...700A..20M}, it is still difficult to predict the outcome of  self-consistent  simulations accurately \cite[e.g.][]{2012ApJ...757...69U,2020MNRAS.491.2715B,2020ApJ...890..127C,2022ApJ...937L..15T}. From this viewpoint, the simple argument proposed in  \citetalias{2025PhRvL.134o1002C} should be verified by performing simulations. 

 \begin{figure}
  \centering
  \includegraphics[width=8cm]{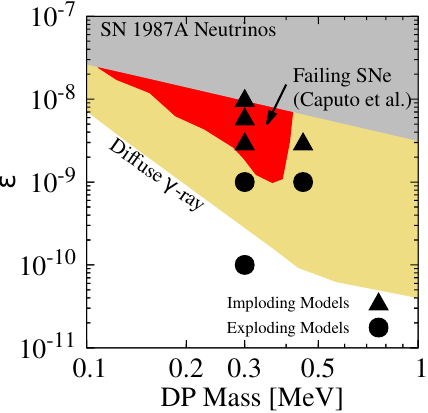}
  \caption{Our seven models are shown on the parameter space for DPs. The circles indicate the exploding models and the triangles indicate the imploding models. The upper limits  on $\epsilon$ \citepalias{2025PhRvL.134o1002C}, which are obtained from the SN 1987A neutrino burst, failing supernovae (SNe), and diffuse DPs, are also shown as the colored regions.}
  \label{models}
 \end{figure}

In this Letter, we report the results of neutrino-radiation hydrodynamic simulations for a collapsing star coupled with the DP cooling effect. We find that, when $m_{\gamma'}=0.3$--0.45\,MeV, the star fails to explode if the DP mixing parameter, $\epsilon$, is higher than $\sim10^{-9}$. Fig.~\ref{models} shows the parameter space for DPs. The seven points indicate the parameters adopted in our simulations and the red region indicates the failing supernova constraint in \citetalias{2025PhRvL.134o1002C}. One can see that, when $m_{\gamma'}=0.3$\,MeV, our models fail to explode if and only if the DP parameters are in the region excluded by the failing supernova argument. This result supports the  argument at least at this DP mass. However, when $m_{\gamma'}=0.45$\,MeV, the model with $\epsilon=3\times10^{-9}$ fails to explode even though the DP parameter is not excluded by \citetalias{2025PhRvL.134o1002C}. This model encourages one to explore wider DP masses with simulations. 

 \section{Method}
 \label{sec:method}

 We consider DPs that are described by the Lagrangian

 \begin{equation}
    \mathcal{L}=\mathcal{L}_\mathrm{SM}+\frac{1}{2}m_\mathrm{\gamma'}^2 A_{\mu}'A'^{\mu}-\frac{1}{4}F'_{\mu\nu}F'^{\mu\nu}-\frac{\epsilon}{2}F'_{\mu\nu}F^{\mu\nu},
 \end{equation}
 where $\mathcal{L}_\mathrm{SM}$ is the Standard Model Lagrangian and $F'_{\mu\nu}=\partial_\mu A'_\nu-\partial_\nu A'_\mu$ is the field strength for the DP field $A'_\mu$. Although DPs are a $U(1)$ gauge boson, they can acquire mass through the Higgs or the Stueckelberg mechanisms~\cite[e.g.][]{2004IJMPA..19.3265R,2020arXiv200501515F}. 
 
 The mixing term $\epsilon F'_{\mu\nu}F^{\mu\nu}/2$ leads to the DP production in stellar plasma. In a dense medium, the photon-DP interaction is significantly enhanced. We can define the effective mixing parameter \cite{2013PhLB..725..190A,2017JHEP...01..107C}
\begin{equation}
    \epsilon_\mathrm{m}^2=\frac{\epsilon^2}{\left(1-\frac{\mathrm{Re}\,\Pi}{m^2_{\gamma'}}\right)^2+\left(\frac{\mathrm{Im}\,\Pi}{m^2_{\gamma'}}\right)^2},\label{em}
\end{equation}
where $\epsilon$ is the mixing parameter in vacuum and $\Pi$ is the polarization tensor. The real part of $\Pi$ is proportional to $\omega_\mathrm{pl}^2$, where $\omega_\mathrm{pl}$ is the effective photon mass in medium, and its explicit expression is shown in Ref.~\cite{1993PhRvD..48.1478B}. The imaginary part is related with the production and absorption rates of photons. In the local thermal equilibrium, it is written as
\begin{equation}
    \mathrm{Im}\,\Pi=-\omega\left(1-e^{-\frac{\omega}{T}}\right)\Gamma_\mathrm{abs}(\omega,\,T),
\end{equation}
where $\omega$ is the photon energy, $T$ is the temperature, and $\Gamma_\mathrm{abs}(\omega,\,T)$ is the absorption width. In this study, we consider the DP production through the nucleon bremsstrahrung ($n+p\rightarrow n+p+\gamma'$), which is a dominant channel in the stellar core \cite{2017JHEP...01..107C,2025PhRvL.134o1002C}. We adopt an expression for $\Gamma_\mathrm{abs}(\omega,\,T)$ in Ref.~\cite{2017JHEP...01..107C}, which is derived under the soft-radiation approximation \cite{1958PhRv..110..974L,1968PhRv..170.1628N,2016PhRvC..94d5805R}. 

Considering the effective mixing parameter in Eq.~(\ref{em}), the DP production rate per unit volume can be evaluated as \cite{2017JHEP...01..107C,2025PhRvL.134o1002C}
\begin{equation}
    Q=\frac{\epsilon^2m_{\gamma'}^4}{2\pi^2}\int d\omega\frac{\omega^2v}{e^\frac{\omega}{T}-1}\sum_{i=\mathrm{T,\,L}}\frac{g_i|\mathrm{Im}\,\Pi_i|}{(m_\mathrm{\gamma'}^2-\mathrm{Re}\,\Pi_i)^2+|\mathrm{Im}\,\Pi_i|^2},\label{Q}
\end{equation}
where $v$ is the DP velocity and the indices $i=\mathrm{T,\,L}$ correspond to the DP transverse and longitudinal modes, respectively. The spin degeneracy is encoded in the factors $g_\mathrm{T}=2$ and $g_\mathrm{L}=1$. Eqs.~(\ref{em}) and (\ref{Q}) show that a resonance appears and the DP production rate is enhanced if $\mathrm{Im}\,\Pi_i\ll\mathrm{Re}\, \Pi_i$, which is satisfied in the stellar core. We numerically evaluate the integral in Eq.~(\ref{Q})  to obtain the production rate when the resonance condition, $\mathrm{Re}\,\Pi_i=m_{\gamma'}^2$, is not satisfied. However, when the resonance condition is satisfied, $Q$ can be analytically written down as \cite{2025PhRvL.134o1002C}
\begin{equation}
    Q\approx\frac{\epsilon^2m_{\gamma'}^4}{2\pi}\sum_{i=\mathrm{T,\,L}}g_i\left|\frac{\partial\mathrm{Re}\,\Pi_i}{\partial \omega}\right|^{-1}\frac{\omega_\mathrm{res}^2v_\mathrm{res}}{e^\frac{\omega_\mathrm{res}}{T}-1},\label{Qr}
\end{equation}
where $\omega_\mathrm{res}$ is the resonance energy that satisfies $\mathrm{Re}\,\Pi_i=m_{\gamma'}^2$ and $v_\mathrm{res}$ is the DP velocity on resonance, by treating the resonance as a $\delta$-function. One can easily show that the resonance for the longitudinal mode appears when $m_{\gamma'}<\omega_\mathrm{pl}$, while the one for the transverse mode appears when $\sqrt{2/3}m_{\gamma'}<\omega_\mathrm{pl}<m_{\gamma'}$.

We perform axisymmetric neutrino-radiation hydrodynamic simulations of a collapsing star with the \texttt{3DnSNe} code \cite{2016MNRAS.461L.112T}, which solves the neutrino transport with the three-flavor isotropic diffusion source approximation \cite{2009ApJ...698.1174L,2014ApJ...786...83T,2018ApJ...853..170K}. We employ the LS220 nuclear equation of state in Ref.~\cite{1991NuPhA.535..331L}. The initial condition of our simulations is a $18.29M_\odot$ blue supergiant model in Ref.~\cite{2018MNRAS.473L.101U}, which can reproduce observational features of the SN~1987A progenitor. Our simulations are first run in the spherically-symmetric configuration until the post-bounce time $t_\mathrm{pb}=0.01$\,s. After that, the simulations are switched to the axisymmetric configuration and the DP cooling effect is turned on. We evaluate the DP production rate with Eqs.~(\ref{Q}) and (\ref{Qr}) in the simulations and reduce the internal energy per unit volume, $e_\mathrm{int}$, as
\begin{equation}
    \left(\frac{\partial e_\mathrm{int}}{\partial t}\right)_{\gamma'}=-Q.
\end{equation}


\section{Results}\label{sec:results}

\begin{table}[]
\begin{tabular}{cc|ccccc}
$m_{\gamma'}$  & $\epsilon_{10}$& Shock Revival? & $t_{\mathrm{pb},\,2000}$&$E_\mathrm{diag}$ & $M_\mathrm{Ni}$ & $M_\mathrm{PNS}$ \\
$\mathrm{[MeV]}$&&&[ms]&[$10^{51}$\,erg]&[$M_\odot$]&[$M_\odot$]\\

\hline
 & 0 &Yes&428& 0.28&0.062 & 1.73\\
0.3  & 1        &Yes&389&0.24&0.055 &1.71 \\
0.3   & 10        & Yes&478&0.21&0.035 &1.75 \\

0.3   & 30        & No&--&--&-- &--   \\
0.3   & 60        &No&--&--&--&-- \\
0.3& 100        &No&--&--&--&--  \\
0.45   & 10        &Yes&523&0.15&0.017&1.76 \\
0.45& 30      &No&--&--&--&--  \\
 \\
\end{tabular}
\caption{The summary of our models. The first two columns indicate the DP mass $m_{\gamma'}$ and the mixing parameter $\epsilon_{10}=\epsilon/10^{-10}$. The diagnostic explosion energy $E_\mathrm{diag}$, the ejected nickel mass $M_\mathrm{Ni}$, and the proto-neutron star mass $M_\mathrm{PNS}$ are also shown. They are measured at $t_\mathrm{pb}=t_\mathrm{pb,\,2000}$, at which the bounce shock reaches a radius of $2000$\,km.}
\label{table}
\end{table}

We perform eight simulations with the DP mass $m_{\gamma'}=0.3$ and 0.45\,MeV and the mixing parameter $\epsilon_{10}=\epsilon/10^{-10}=0$, 1, 10, 30, 60, and 100 as shown in Fig.~\ref{models}. The model with $\epsilon_{10}=0$ corresponds to the reference model without the DP effect. The results of our simulations are summarized in Table \ref{table}.

\subsection{Models with $m_{\gamma'}=0.3$\,MeV}

  \begin{figure}
  \centering
  \includegraphics[width=8.5cm]{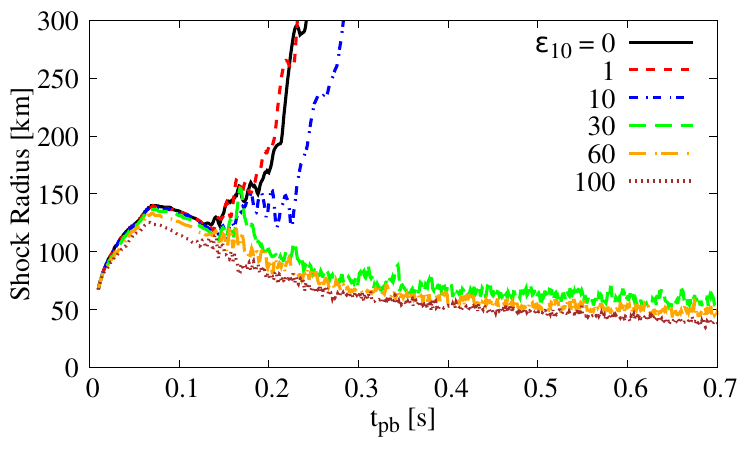}
  \caption{The angular-averaged shock radius as a function of the post-bounce time $t_\mathrm{pb}$. The solid curve corresponds to the reference model without DPs and the other curves correspond to the models with DPs. The DP mass is fixed to  $m_{\gamma'}=0.3$\,MeV.}
  \label{rsh}
 \end{figure}

   \begin{figure}
  \centering
  \includegraphics[width=8.5cm]{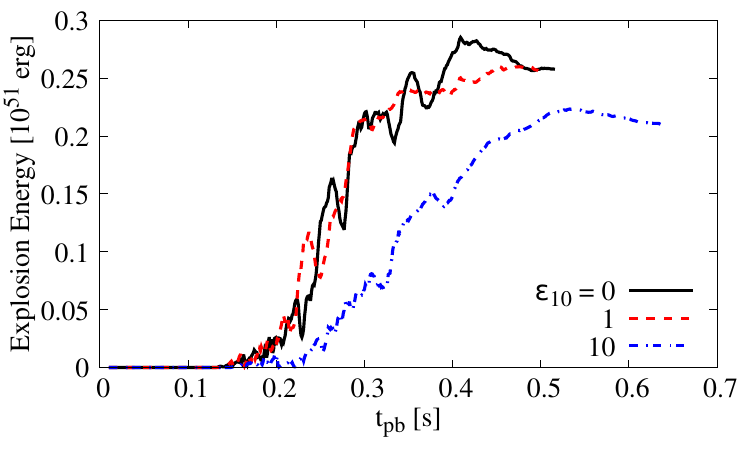}
  \caption{The diagnostic explosion energy $E_\mathrm{diag}$ for the exploding models as a function of the post-bounce time $t_\mathrm{pb}$. The solid curve corresponds to the reference model without DPs and the other curves correspond to the models with DPs. The DP mass is fixed to  $m_{\gamma'}=0.3$\,MeV.}
  \label{Eexp}
 \end{figure}

We first describe the models with $m_{\gamma'}=0.3$\,MeV. Figure \ref{rsh} shows the radius of the bounce shock of our models. In the reference model with $\epsilon_{10}=0$, the bounce shock is stalled in the early phase at $t_\mathrm{pb}\lesssim0.15$\,s and it propagates outward in the later phase. This behavior of the bounce shock is typical for delayed supernova explosions confirmed by recent multi-dimensional simulations \citep[e.g.][]{2011ApJ...738..165S,2014ApJ...786...83T,2015ApJ...807L..31L,2015ApJ...808L..42M,2015ApJ...801L..24M,2016ApJ...817...72P,2019JPhG...46a4001P,2017MNRAS.472..491M,2018ApJ...865...81O,2018ApJ...855L...3O,2020MNRAS.491.2715B,2024ApJ...964L..16B,2020ApJ...896..102K,2021ApJ...915...28B,2025MNRAS.536..280N}. As shown in the figure, when $\epsilon_{10}\leq10$, our models show the successful shock revival. The models with $\epsilon_{10}=0$ and 1 show the shock revival almost at the same time at $t_\mathrm{pb}\sim0.15$\,s. When $\epsilon_{10}=10$, the shock revival is delayed until $t_\mathrm{pb}\sim0.23$\,s. On the other hand, when $\epsilon_{10}\geq30$, the situation is qualitatively different. The shock revival does not happen until the end of the simulations in these models. In this case, the mass accretion on the proto-neutron star continues and finally a black hole will be formed at the center. 

The models without the shock revival could be observed as a failed supernova, one of whose candidates is M31-2014-DS1 \cite{2024arXiv241014778D,2025arXiv250419510S,2025arXiv251103470N}. However, our simulations adopt the progenitor model which is fine-tuned for  SN~1987A~\cite{2018MNRAS.473L.101U}, which obviously exploded. The comparison between the SN~1987A explosion and our models provides a constraint, $\epsilon<3\times10^{-9}$, at $m_{\gamma'}=0.3\,$MeV.  This result is close to the upper limit $\epsilon\lesssim2\times10^{-9}$ reported by \citetalias{2025PhRvL.134o1002C} with the post-processing technique. 
A systematic exploration of the progenitor dependence \cite{2016ApJ...825....6S,2020MNRAS.491.2715B,2025MNRAS.536..280N} lies beyond the scope of this study and is left for future investigations.

Figure~\ref{Eexp} shows the diagnostic explosion energy,
 \begin{eqnarray}
    E_\mathrm{diag}=\int_D dV\left(\frac{1}{2}\rho v^2+e-\rho\Phi\right),\label{Ediag}
\end{eqnarray}
for the exploding models. Here $\rho$ is density, $v$ is the fluid velocity, $e$ is the internal energy density ,$\Phi$ is the gravitational potential, and $D$ is the region where the integrand is positive. One can find that $E_\mathrm{diag}$ saturates at $\sim0.25\times10^{51}$\,erg when $\epsilon_{10}\leq1$. When $\epsilon_{10}$ is as high as $10$, the explosion energy is reduced to $\sim0.21\times10^{51}$\,erg. Although the number of our exploding models is limited, this result hints that DPs can weaken supernova explosions even if the mixing parameter is not excluded by the failing supernova argument.

Table \ref{table} shows the ejected nickel mass $M_\mathrm{Ni}$ and the proto-neutron star mass $M_\mathrm{PNS}$ for our models. One can find that $E_\mathrm{diag}$ and $M_\mathrm{Ni}$ are reduced by the DP cooling. This is similar to previous two-dimensional models coupled with the production of light sterile neutrinos \cite{2025PhRvD.111h3046M}. On the other hand, the values for $M_\mathrm{PNS}$ coincide with each other within a few percents. This looks counterintuitive because the additional cooling is expected to enhance the mass accretion rate and thus $M_\mathrm{PNS}$. The reason for the small difference in $M_\mathrm{PNS}$ could be attributed to stochasticity, which is often reported in multi-dimensional supernova models. It is desirable to develop additional supernova models with $\epsilon_{10}=10$--30 to investigate the DP effect on $M_\mathrm{PNS}$, but the task is beyond the scope of this study. 

   \begin{figure}
  \centering
  \includegraphics[width=8.5cm]{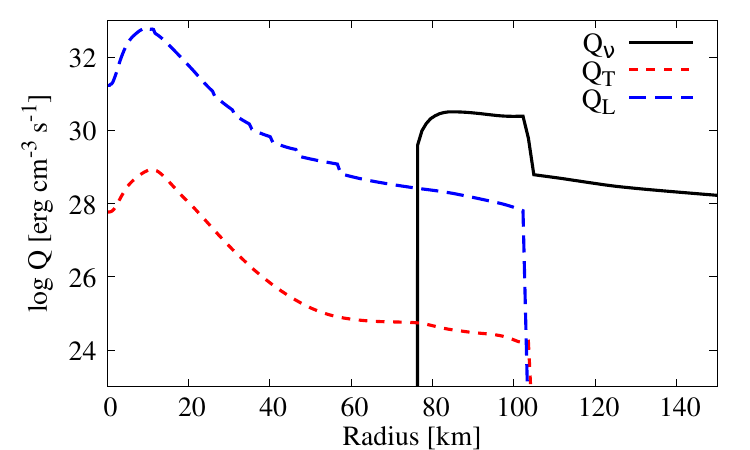}
  \includegraphics[width=8.5cm]{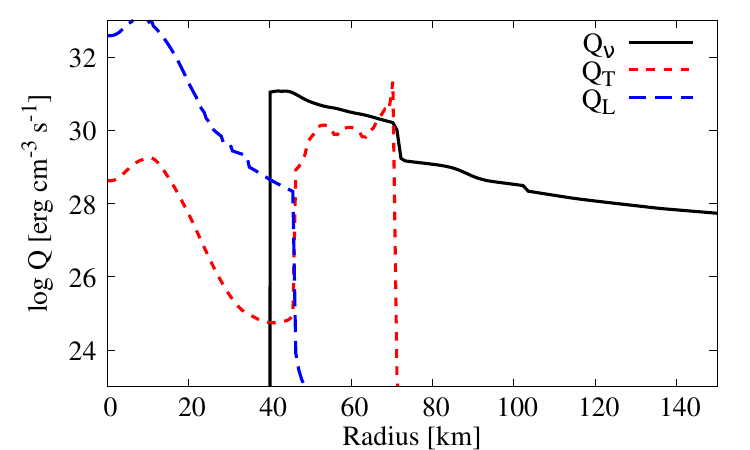}
  \caption{The radial profile of the neutrino heating rate $Q_\nu$ and the transverse (longitudinal) DP cooling rate $Q_\mathrm{T}$ ($Q_\mathrm{L})$ in the equatorial plane for the $m_{\gamma'}=0.3$\,MeV and $\epsilon_{10}=30$ model. The upper panel is the snapshot at  $t_\mathrm{pb}=0.15$\,s and the lower panel is the snapshot at  $t_\mathrm{pb}=0.30$\,s.}
  \label{Q_DP}
 \end{figure}

     \begin{figure}
  \centering
  \includegraphics[width=8.5cm]{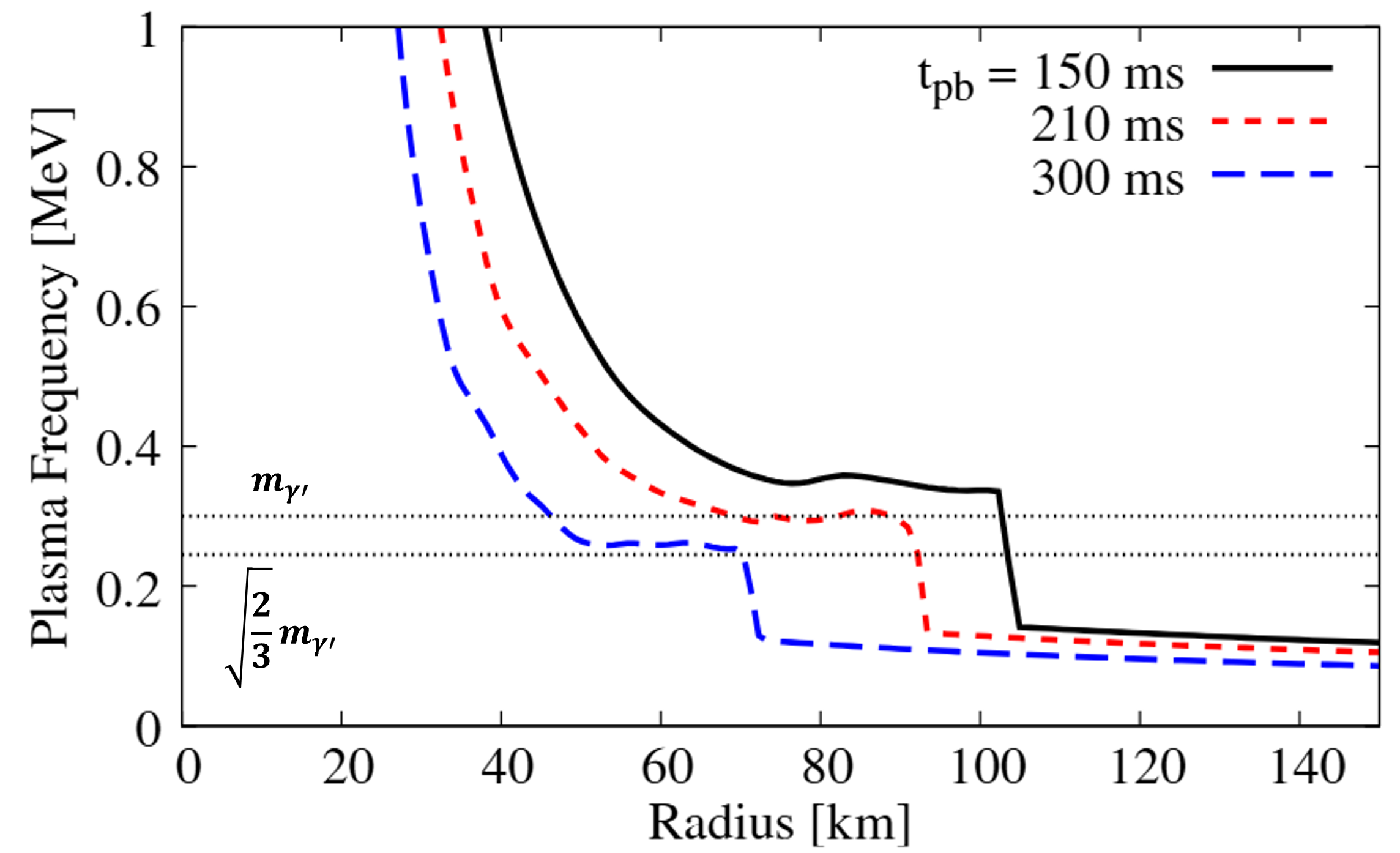}
  \caption{The radial profile for the plasma frequency $\omega_\mathrm{pl}$ in the equatorial plane for the $\epsilon_{10}=30$ model. The horizontal lines show the DP mass $m_{\gamma'}=0.3$\,MeV and $\sqrt{2/3}m_{\gamma'}$, which indicate the region where DPs are resonantly produced.}
  \label{plasma}
 \end{figure}

    \begin{figure}
  \centering
  \includegraphics[width=8.5cm]{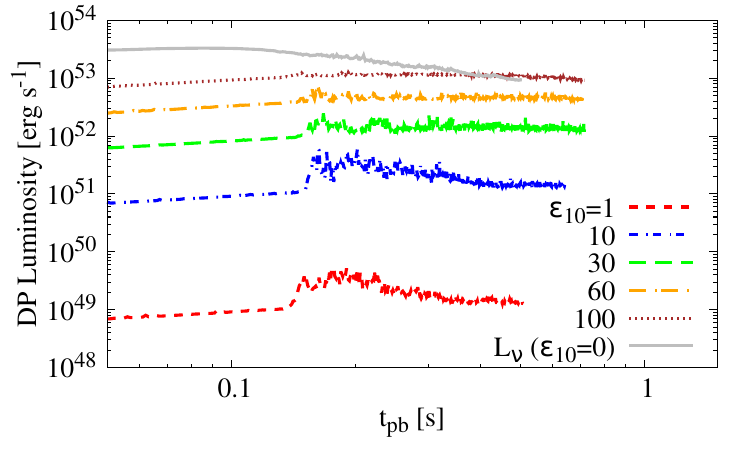}
  \caption{The DP luminosity as a function of the post-bounce time $t_\mathrm{pb}$. The total neutrino luminosity in the $\epsilon_{10}=0$ model is also shown by the solid gray curve. The DP mass is fixed to  $m_{\gamma'}=0.3$\,MeV.}
  \label{L_DP}
 \end{figure}

The hindrance of supernova explosions is caused by the DP production in the gain region. Figure \ref{Q_DP} shows the radial profile of the neutrino heating rate and the DP cooling rate in the $\epsilon_{10}=30$ model. In an early phase at $t_\mathrm{pb}\lesssim0.20$\,ms, the dominant process for the DP production is the resonant production of longitudinal-mode DPs. The total DP cooling rate, $Q=Q_\mathrm{T}+Q_\mathrm{L}$, is 2--3 orders of magnitude smaller than the neutrino heating rate, $Q_\nu$. In the later phase at $t_\mathrm{pb}\gtrsim0.20$\,ms, the resonant production of transverse DPs becomes dominant in the gain region.  In this phase, the DP cooling rate is as large as $Q_\nu$. In the standard paradigm of the neutrino-driven explosion mechanism, this neutrino heating is essential to unbind the accreting matter and make successful explosions \cite{1985ApJ...295...14B}. The strong DP cooling counterbalances the neutrino heating and hinders the explosion, as shown in Fig.~\ref{rsh}.

The appearance of the transverse and longitudinal-mode resonances can be understood from the radial profile of the plasma frequency, which is shown in Fig.~\ref{plasma}. One can prove that the transverse-mode resonance appears when $m_{\gamma'}<\omega_\mathrm{pl}$ and the longitudinal-mode resonance appears when $\sqrt{2/3}m_{\gamma'}<\omega_\mathrm{pl}<m_{\gamma'}$. The figure shows that, at $t_\mathrm{pb}=0.15$\,s, $\omega_\mathrm{pl}$ in the gain region exceeds $m_{\gamma'}$, and as a result, the transverse-mode resonance appears. The plasma frequency gradually decreases as the time goes on and it becomes smaller than $m_{\gamma'}$ at $t_{\mathrm{pb}}\gtrsim0.20$\,s. In this phase, the condition for the longitudinal-mode resonance is satisfied. 

Figure \ref{L_DP} shows the DP luminosity $L=\int QdV$ as a function of time. One can see that $L$ is proportional to $\epsilon_{10}^2$ at $t_{\mathrm{pb}}\lesssim0.1$\,s because the hydrodynamic profiles in each model do not significantly deviate from each other. At $t_\mathrm{pb}\gtrsim0.15$--0.20\,s, the transverse-mode resonance appears and $L$ is enhanced, although its contribution depends on the models. In the model with $\epsilon_{10}=100$, the DP luminosity exceeds the total neutrino luminosity at $t_\mathrm{pb}\sim0.4$\,s. In such a case, the neutrino luminosity will be significantly reduced and the mixing parameter is excluded by the so-called energy-loss argument \cite{2009PhRvD..80g5018B,2016PhRvC..94d5805R,2017JHEP...02..033H,2017JHEP...01..107C}.

\subsection{Models with $m_{\gamma'}=0.45$\,MeV}

  \begin{figure}
  \centering
  \includegraphics[width=8.5cm]{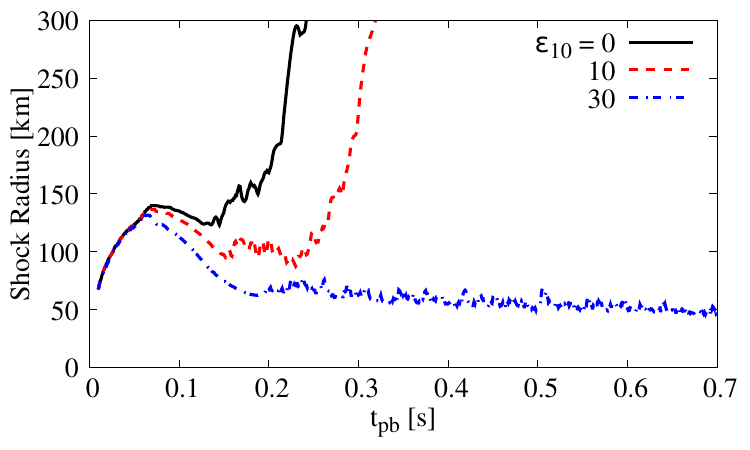}
  \caption{The same plot as Fig.~\ref{rsh} but the DP mass is fixed to  $m_{\gamma'}=0.45$\,MeV.}
  \label{rsh_0.45}
 \end{figure}

Figure \ref{rsh_0.45} shows the angular-averaged shock radius for the models with $m_{\gamma'}=0.45$\,MeV. We can see that the stalled shock is revived in the $\epsilon_{10}=10$ model. On the other hand, the shock revival does not happen in the model with $\epsilon_{10}=30$. This result suggests that we can obtain an upper limit $\epsilon_{10}<30$ on the mixing parameter at $m_{\gamma'}=0.45$\,MeV based on the failing supernova argument. Interestingly, as shown in Fig.~\ref{models}, the mixing parameter $\epsilon_{10}\sim30$ is not excluded at this DP mass by \citetalias{2025PhRvL.134o1002C}, which is based on the post-processing. This difference highlights the importance of self-consistent simulations, which can calibrate the constraint obtained by the post-processing.


\section{SUMMARY and Discussion}\label{sec:sum-dis}

In this study, we performed two-dimensional neutrino-radiation hydrodynamic simulations for a collapsing star coupled with the production of DPs with $m_{\gamma'}=0.3$ and 0.45\,MeV. As the photon-DP mixing parameter is increased, the successful supernova explosion transitioned to failures in the explosion. This suggests that the mixing parameter can be constrained by the very existence of supernova explosions in the Universe, as proposed by \citet[][C25]{2025PhRvL.134o1002C}. Our models show quantitative agreement with the constraint obtained in \citetalias{2025PhRvL.134o1002C} in the case of $m_{\gamma'}=0.3$\,MeV, encouraging to extend the argument to various exotic particle models. 

Supernova explodability  is dependent on the spatial dimension of simulations. A way to quantify explodability is the critical neutrino luminosity  as a function of the mass accretion rate, above which steady state solutions do not exist \cite{1993ApJ...416L..75B,2005ApJ...623.1000Y}. Multi-dimensional hydrodynamic simulations have revealed that the critical luminosity is the highest for one-dimensional models and the lowest for two-dimensional models, and the one for three-dimensional models is between the two \cite{2008ApJ...688.1159M,2012ApJ...755..138H,2012ApJ...749...98T,2013ApJ...775...35C}. This suggests that  three-dimensional models may explode even if we consider higher values of the photon-DP mixing parameter. In this study, we developed two-dimensional models to reduce the computational cost. Our constraint is conservative in a sense that the models are easier to explode than three-dimensional models. It is still desirable to perform three-dimensional simulations coupled with DPs in the future, because they are expected to lead to a more stringent constraint.

It is also important to extend this study to investigate a wider DP mass range. In our simulations, the DP mass is fixed to $m_{\gamma'}=0.3$ and 0.45\,MeV. However, the failing supernova constraint is stronger than the cooling constraint for DPs with $m_{\gamma'}\sim0.1$--0.4\,MeV \cite{2025PhRvL.134o1002C}. Although our results showed that simulations and post-processing lead to a similar result when $m_{\gamma'}=0.3$\,MeV, it is not guaranteed that simulations can reproduce the previous result for other DP masses. In fact, our simulations suggest that, when $m_{\gamma'}=0.45$\,MeV, the mixing parameter should satisfy $\epsilon<3\times10^{-9}$  to obtain successful supernova explosions, while the parameter is located in the allowed region based on the post-processing. In addition, DPs heavier than $\sim1$\,MeV can decay into a electron-positron pair. In this case, DPs can contribute to the energy transfer in a star and enhance the explosion energy. In our simulations, we did not consider the DP heating rate because the DP mean free path is long enough. However, when DPs are unstable or the mixing parameter is much higher, hydrodynamic simulations that solve the DP transfer consistently should be performed, as done for other exotic particles such as axions \cite{2022JCAP...08..045C,2022PhRvD.105f3009M,2023PhRvD.108f3027M,2025PhRvD.111j3028T} and sterile neutrinos \cite{2014PhRvD..90j3007W,2018PhRvD..98j3010R,2024PhRvD.110b3031M}.

Last but not least, it is worthwhile to perform simulations coupled with DPs for a wide range of progenitors to investigate the DP effect on the mass function of neutron stars and black holes. Surveying the vast parameter space with multi-dimensional simulations will require a  large amount of computational resources. To circumvent this problem, one may perform the so-called 1D+ simulations, which take turbulence into account  phenomenologically \cite{2019ApJ...887...43M,2019MNRAS.487.5304M,2020ApJ...890..127C,2021ApJ...912...29B,2022ApJ...926..147B,2024MNRAS.528.1158S}. 

\begin{acknowledgments}
Numerical computations were  carried out on Cray XD2000 at Center for Computational Astrophysics, National Astronomical Observatory of Japan. This work was in part supported by JSPS KAKENHI Grant Numbers K.M.: JP23KJ2147, JP23K13107, JP25H02194;
T.T.: JP23K25895, JP23K22494, JP23K03400, JP24K00631;
K.K.: JP23KF0289, JP24K07027 and MEXT KAKENHI Grants Number JP24H01825 (K.K.).
This work was also supported by “Program for Promoting researches on the Supercomputer Fugaku” (Structure and Evolution of the Universe Unraveled by Fusion of Simulation and AI; Grant Number JPMXP1020230406), and JICFuS.
\end{acknowledgments}

\bibliography{ref.bib,refarXiv}
\end{document}